\documentstyle[aps,prl,psfig]{revtex}
\draft
\begin{document}
\twocolumn[\hsize\textwidth\columnwidth\hsize\csname
@twocolumnfalse\endcsname
\title{Entropic Elasticity, Cooperative Extensibility and
Supercoiling Property of DNA: A Unified Viewpoint}
\author{Zhou Haijun\cite{zhouhj}$^{1,2}$,
Zhang Yang$^1$, and Ou-Yang Zhong-can$^{1,3}$}
\address{$^{1}$Institute of Theoretical Physics, 
Academia Sinica, P.O. Box 2735, Beijing 100080, China\\
$^{2}$State Key Laboratory of Scientific 
and Engineering Computing, Beijing 100080, China\\
$^3$Center for Advanced Study, Tsinghua University, Beijing
100084, China}

\date{27 January, 1999 (PRL copyright)}
\maketitle

\begin{abstract}

A unified  model is constructed to study 
the recently observed  DNA entropic elasticity, 
 cooperative extensibility, and supercoiling property. 
With the introduction of a new structural parameter 
(the folding angle $\varphi$), bending deformations
of  sugar-phosphate backbones, steric effects of 
nucleotide basepairs, and short-range basestacking 
interactions are considered.  The comprehensive agreement of 
theoretical results with experimental observations on both 
torsionally relaxed  and negatively supercoiled DNAs 
 strongly indicates that,  basestacking 
interactions, although   short-ranged in nature, dominate 
the elasticity of DNA and hence are  of vital biological 
significance.  

\end{abstract}

\vskip2pc]

Recent {\it in vitro} 
experiments done
on single double-stranded DNA (dsDNA) molecules or 
 DNA-protein complexes reveal that  DNA double-helix
has nontrivial elastical properties 
\cite{smith92,west92,cluzel96,smith96,strick96,strick98,allemand98}.
At  low external forces, it
can be viewed as a simple  wormlike chain \cite{marko95a};
at moderate forces, it becomes a  rod with a large stretch modulus.
But if  pulled with a
 large force  of about $70$ pN 
the molecule as a whole can suddenly be  driven
  to an almost fully stretched state with a contour
length $1.7$ times its native value \cite{cluzel96,smith96}.
More strikingly, if a dsDNA  at the same time
has a slight deficit in linking number,
 {\it i.e.}, negatively supercoiled,
a pulling force as small as
 $0.3$ pN can distort the
native structure of  DNA considerablely 
\cite{strick96,strick98,remark3}.

On the theoretical side, to
understand these novel properties of DNA is of current 
interest and  many models are
suggested  and some valuable insights   attained
\cite{marko95a,marko95b,fain97,marko97,cizeau97,rief98,bouchiat98,kamien97,ha97,kroy96,ahsan98,zhou99}. 
For example, to interpret the extensibility of DNA,
some authors  suggest one-dimensional two-state
models with \cite{cluzel96,marko97,cizeau97,ahsan98} or without
\cite{rief98} nearest-neighbor interactions;
and to explain the supercoiling property of DNA,
 wormlike rod chain models, with \cite{marko97,kamien97}
or without \cite{marko95b,fain97,bouchiat98}
  bend-twist and/or
 stretch-twist coupling,  are investigated. Nevertheless,
 a unified description still seems to be  lacking
and the underlying mechanism
which should account for 
DNA cooperative extensibility and 
novel supercoiling properties 
still elusive.   
Here, we show that it is possible to
 understand all these experimental observations
from a unified viewpoint.

A simple elastical model
 is proposed by taking into account 
the  structural properties
of realistic dsDNA. Bending energy 
of the sugar-phosphate backbones, 
basestacking interaction  
between adjacent nucleotide basepairs as well
as their steric effects   on DNA axial bending 
rigidity  are considered. We have 
introduced a new structural
parameter,  the folding angle $\varphi$.
Model calculations on the elastical properties of
torsionally relaxed and negatively supercoiled DNAs
are in quantitative agreement with  all the known
 experimental
observations \cite{smith92,cluzel96,smith96,strick96,strick98}.
The model indicates 
that basestacking interaction is
 the main factor determining the 
highly extensibility and unwinding instability of DNA. 
We suggest that the present model, 
after some revisions, will also be 
able to account for the elasticity of 
positively supercoiled DNAs \cite{strick96,allemand98}.  

In the model, the two inextensible  backbones of DNA
\cite{saenger84} are characterized
by the same bending rigidity $\kappa=k_B T \ell_p$, where $\ell_p\simeq 1.5$ nm
is  their bending persistence length (BPL).
  Their position
vectors are  ${\bf r}_i=\int^{s} 
{\bf t}_i(s^\prime)d s^\prime$, where
 ${\bf t}_i$ ($i=1,2$) is the unit tangential vector of the $i$th backbone, and $s$
 its arc length. The nucleotide basepairs 
between the backbones \cite{saenger84}
are viewed as rigid planar structures 
with finite area and volume.  
Firstly, we consider  bending energy of the backbones 
alone and each basepair connecting the two
backbones  is replaced  for the moment
by a thin rigid rod of  length  $2 R$, with  a unit vector ${\bf b}$
 pointing along it from ${\bf r}_1$ to ${\bf r}_2$, {\it i.e.},
 ${\bf r}_2(s)-{\bf r}_1(s)=2 R{\bf b}(s)$. 
Relative sliding of the 
backbones is prohibited, the basepair planes are assumed to lie perpendicular
to  DNA central axis  
and ${\bf b}\cdot{\bf t}_1={\bf b}\cdot {\bf t}_2\equiv 0$
\cite{liverpool98}. 
The central axis of dsDNA can be defined as ${\bf r}(s)={\bf r}_1(s)+R {\bf b}(s)$, and its
tangential vector is denoted by ${\bf t}$, 
with ${\bf t}\cdot {\bf b}= 0$.
Since both ${\bf t}_1$
 and ${\bf t}_2$ lie on the
 same plane perpendicular to ${\bf b}$,
we obtain  that
${\bf t}_1=
{\bf t}\cos\varphi+{\bf n}\sin\varphi$ and ${\bf t}_2=
{\bf t}\cos\varphi-
{\bf n}\sin\varphi$,
 where ${\bf n}={\bf b}\times {\bf t}$ and
$\varphi$ is half the rotational angle from ${\bf t}_2$ to ${\bf t}_1$ ({\bf b} being
the rotational axis). We call $\varphi$  the folding angle, it is  in the range
between $-\pi/2$ and $+\pi/2$
($\varphi >0$ for right-handed rotations and $<0$ for
left-handed ones).
It is not difficult to verify that
\begin{equation}
{d{\bf b}/ ds}={({\bf t}_2-{\bf t}_1)/2 R}=-
{\bf n} \sin\varphi/R,
\label{eq1}
\end{equation}
(here and after  $d s$   always  denotes  arc 
length element of the  {\it backbones}). 
With Eq. (\ref{eq1}) and the definition 
of ${\bf r}$ we know that
\begin{equation}
{d{\bf r}/ds}={({\bf t}_1+{\bf t}_2)/2}={\bf t}\cos\varphi.
\label{eq2}
\end{equation}
Then total bending energy of the backbones, 
$E_{b}=\int (\kappa/2)[(d{\bf t}_1 /ds)^2
+(d{\bf t}_2/ds)^2] ds$ \cite{liverpool98},
 can be rewritten,
 with the help of Eqs. (\ref{eq1}) and (\ref{eq2}), as
\begin{equation}
E_b=\int_0^L\left[\kappa({d{\bf t}\over ds})^2+\kappa({d\varphi\over ds})^2
+{\kappa\over R^2}\sin^4\varphi\right]ds,
\label{eq3}
\end{equation}
 here $L$ is the total contour length
of each backbone.
This expression proves to be very useful.
 The  second and the third terms in Eq.(\ref{eq3}) is deformation energy
caused by folding of the backbones with respect to the central axis, and
the first term, $\kappa (d{\bf t}/ds)^2$, is the bending energy of  DNA 
central axis contributed by the backbone 
bending rigidity $\kappa$. So far, basepairs  are
 viewed as  thin rods and their contribution to the bending 
rigidity of  DNA chain is not considered. 
 Because of steric effects caused by finite volume and area,
 basepairs 
will certainly increase the bending rigidity of  
DNA chain \cite{remark1}.  
 The simplest way to consider such  effects  is
to replace $\kappa$ in the first term 
of Eq. (\ref{eq3}) with a
 phenomenological parameter $\kappa^*$, with $\kappa^* > \kappa$.
Hereafter this  is assumed.

Besides steric effects, nucleotide 
basepairs contribute also basestacking energy.
This energy  mainly originates from noncovalent
  van der Waals interactions
between adjacent basepairs \cite{saenger84}.
Basestacking interaction is
 short-ranged and  is characterized by an 
attraction potential
proportional to $1/r^6$ and a strong 
repulsion potential proportional to $1/r^{12}$ (here
$r$ is the axial distance between adjacent basepairs). 
In our continuous model, the line density 
of such Lennare-Jones type potential
can be  written as 
\begin{equation}
\rho(\varphi)=
\left\{\begin{array}{lll}
{\epsilon\over r_0}[({\cos\varphi_0\over\cos\varphi})^{12}
-2({\cos\varphi_0\over\cos\varphi})^6] &\;\;{\rm for}&
\;\;(\varphi\geq 0),\\
{\epsilon\over r_0}[\cos^{12}\varphi_0-
2\cos^6\varphi_0]&\;\;{\rm for}&\;\; 
(\varphi <0), 
\end{array} \right.
\label{eq4}
\end{equation}
and the total basestacking energy is $E_{LJ}=\int_0^L \rho ds$. 
In Eq. (\ref{eq4}), $r_0$ is the backbone arclength
between adjacent bases; $\varphi_0$ is a parameter related to the
equilibrium distance between a DNA dimer; $\epsilon$ is the basestacking
intensity which  is  generally 
base-sequence specific. Here we focus on 
 macroscopic properties of DNA and  just consider $\epsilon$ in
the average sense and take it as a constant, 
with $\epsilon\simeq 14.0 k_B T$
as indicated by quantum chemical calculations \cite{saenger84}. 
The asymmetric basestacking potential Eq. (\ref{eq4})
 ensures a relaxed  DNA
 to take on a    
right-handed double-helix configuration with its folding angle $\varphi\sim
\varphi_0$.
However, if adjacent basepairs are pulled apart slightly from the equilibrium
distance by external forces or thermal stretching fluctuations, the
basestacking interaction
intensity quickly decreases because of its 
 short-range nature.  
 In other words, the basestacking potential
 can  endure  only a limited pulling  force.
 We believe this to be closely 
related to the observed 
DNA highly cooperative extensibility. 
It may also account for 
 the novel elasticity of
negatively supercoiled dsDNA, since negative supercoiling actually  leads to
an effective pulling force.
This insight, which is developed in more detail in the
following, seems to
be confirmed by experiments \cite{remark4}.

We first discuss the elastic  response of the
 model DNA   when 
a pulling force ${\bf F}=f{\bf z}_0$ along
direction ${\bf z}_0$  is applied at its end.
The total energy functional  is then
$E=E_b+E_{LJ}-\int_0^L f\cos\varphi {\bf t}\cdot {\bf z}_0 ds$. And
the Green function $G({\bf t},\varphi;{\bf t}^\prime,\varphi^\prime;s)$
\cite{marko95a,doi86}, 
 which determines the probability distribution
of ${\bf t}$ and $\varphi$ along DNA chain, is governed by
\begin{equation}
{\partial G\over \partial s}=\left[{\partial^2 
\over 4 \ell_p^* \partial {\bf t}^2}
+{\partial^2\over 4\ell_p \partial \varphi^2}
-{f\over k_B T}cos\varphi 
{\bf t}\cdot {\bf z}_0 -V(\varphi)\right]G,
\label{eq5}
\end{equation}
where  $\ell_p^*=\kappa^*/k_B T$ and $V(\varphi)=\rho(\psi)/k_B T+
\ell_p \sin^4\varphi/R^2$. The spectrum of the above Green  equation
is discrete and hence for  long chains, the average extension
can be obtained either by differentiation of the ground-state eigenvalue, $g$,
of Eq. (\ref{eq5})  with respect to $f$:
\begin{equation}
{\langle Z\rangle/L}=
{(1/L)}\int_0^L \langle
 cos\varphi {\bf t}\cdot {\bf z}_0 \rangle ds
=k_B T {\partial g/\partial f},
\label{eq6}
\end{equation}
or by a direct integration with  the normalized
 ground-state eigenfunction, 
$\Phi({\bf t},\varphi)$,
 of Eq. (\ref{eq5}):
\begin{equation}
{\langle Z\rangle}/L=\int\int 
 |\Phi|^2 {\bf t}\cdot {\bf z}_0\cos\varphi 
d{\bf t} d \varphi.
\label{eq7}
\end{equation}

\vspace*{1.0cm}
\begin{figure}
\centerline{\psfig{file=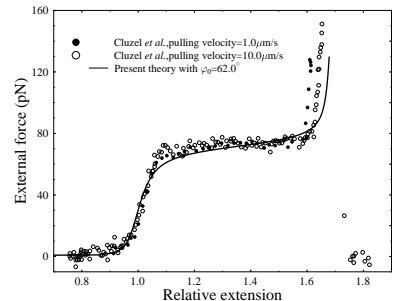,height=5cm}}
\vspace*{-1.0cm}\caption{
Force-extension relation of DNA.
 Experimental data is from
Fig. 2A of [3]. Theoretical curve is
obtained by the following considerations: (i) 
$\ell_p=1.5$ nm [4] and $\epsilon=14.0 k_B T$
[21]; (ii) $\ell_p^*=53.0/2\langle \cos
\varphi\rangle_{f=0}$ nm [30], 
$r_0=0.34/\langle\cos\varphi\rangle_{f=0}$ nm and
$R=(0.34\times 10.5/2\pi)\langle \tan\varphi\rangle_{f=0}$
nm [31];
(iii) adjust the value of $\varphi_0$ to fit the data.
For each $\varphi_0$, the value of $\langle\cos\varphi\rangle
_{f=0}$ is obtained self-consistently. 
The present curve is drawn with
 $\varphi_0=62.0^{\circ}$ (in close consistence with the
 structural property of DNA [21]),  
 and $\langle\cos\varphi\rangle_{f=0}$ is
determined to be $0.573840$. DNA extension is scaled with
 its {\bf B}-form contour length $L\langle\cos\varphi\rangle_{f=0}$.
}
\label{fig1}
\end{figure}

\begin{figure}
\centerline{\psfig{file=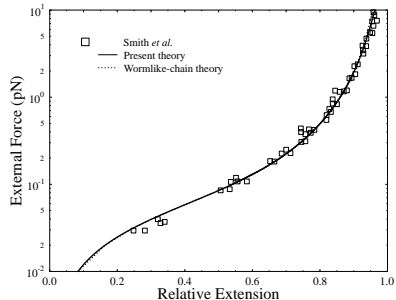,height=5cm}} 
\vspace*{-1.0cm}\caption{
Low-force  elastical behavior of DNA. Here
experimental data is from Fig. 5B of [1], 
the dotted curve is obtained for a wormlike chain with
BPL $53.0$ nm and the parameters for the 
slid curve are the same as those in Fig. \ref{fig1}.
}
\label{fig2}
\end{figure}

Both $g$ and $\Phi({\bf t},\varphi)$ can be 
 obtained numerically through standard diagonalization methods and 
identical results are obtained by Eqs. (\ref{eq6}) and (\ref{eq7}). The resulting
force vs extension relation in the whole relevant force range is shown in
Fig. \ref{fig1} and Fig. \ref{fig2}.
Our theoretical curves are obtained with just
 one adjustable
parameter (see caption of Fig. \ref{fig1}), the
 agreement with  experiments  is strikingly excellent.
According to our theory, 
 the onset of cooperative extension 
of DNA axial length at
forces about $70$ pN is  mainly 
caused by  the 
yielding of  the short-range basestacking interaction
\cite{remark2}.

Below the onset of cooperative elongation, DNA seems to be very stiff and
calculations show that at $f=50$ pN the total extension of DNA is only 
$4.1\%$ longer than its {\bf B}-form 
contour length, in close accordance with 
the value of 
$4.6\%$ reported by Smith {\it et al.} \cite{smith96}.
This is related to the 
fact that the basestacking intensity $\epsilon$ is  
very strong \cite{remark2}.
 At low forces ($f< 10$ pN), because the fluctuation
of the folding angle $\varphi$ is extremely small, it can just 
be neglected and   DNA elasticity is caused by thermal fluctuations
of the axial direction ${\bf t}$ (entropic elasticity).
It is easy to prove that the 
now well-known entropic elasticity (wormlike chain)
 model \cite{marko95a} with
contour length $L\langle \cos\varphi\rangle_{f=0} $
and persistence length $2\ell_p^*\langle 
\cos\varphi\rangle_{f=0}$
 is just an excellent approximation of the present theory (here $\langle
\cos\varphi\rangle_{f=0}$ is the average of $\cos\varphi$ at zero force).
  This point is
demonstrated clearly in Fig. \ref{fig2}.

\vspace*{1.0cm}

Now we continue to study the supercoiling property of the
 model DNA.
Mathematically, a supercoiled dsDNA is characterized by
its fixed value of linking  number $Lk$. It measures the total topological
turns one DNA backbone winds around the other or around the central axis, 
and can be expressed as the sum of the twisting number, $Tw({\bf r}_1,{\bf r})$,
of backbone ${\bf r}_1$ around the central axis ${\bf r}$ and the writhering
number $Wr({\bf r})$ of the central axis,
 {\it i.e.}, $Lk=Tw+Wr$ \cite{white69,fuller78}.
According to Eq. (\ref{eq1}), $Tw({\bf r}_1,
{\bf r})=\int {\bf t}\times (-{\bf b})\cdot d (-{\bf b})=
\int \sin\varphi ds/R$ \cite{fuller78,white69}. For a linear  DNA chain,
the writhering number of its central axis
 can be expressed as \cite{fuller78}
\begin{equation}
Wr({\bf r})=\int {{\bf z}_0\times {\bf t} \cdot d({\bf z}_0+{\bf t})/ds
\over 1+{\bf z}_0\cdot {\bf t} } ds .
\label{eq8}
\end{equation}
The elasticity of such a supercoiled DNA 
chain is determined by the following
energy functional: $E=E_b+E_{LJ}-f\int\cos\varphi 
{\bf t}\cdot {\bf z}_0 ds
-\Gamma k_B T Lk$, where $\Gamma k_B T$ is  torque associated with the
topological constraint.
However, the writhering number expression given by
Eq.(\ref{eq8}) is correct only for 
${\bf t}\cdot {\bf z}_0 \neq -1$,
{\it i.e.}, for chains whose tangential vector ${\bf t}$ never points 
to $-{\bf z}_0$ \cite{fuller78}.
 This condition is satisfied
actually  only for  a highly extended
chain whose ${\bf t}$ fluctuates slightly around ${\bf z}_0$. 
In this case  Eq. (\ref{eq8}) leads to
 $Wr({\bf r})\simeq
(1/2)\int (t_x d t_y/ds-t_y dt_x/ds) ds$,
 where $t_x$ and $t_y$ are 
respectively the $x$ and $y$ component 
of ${\bf t}$. This
approximation is used hereafter.
 If we are to use Eq.(\ref{eq8}) in the general case,
a cutoff procedure seems necessary to
 avoid divergent results \cite{bouchiat98}.

 The  Green equation
for this case is written as
\begin{equation}
\begin{array}{ll}
{\partial G\over \partial s}=&[{\partial^2\over
4\ell_p^*\partial {\bf t}^2}+{\partial^2 \over
4\ell_p \partial \varphi^2}+
{f\cos\varphi\over k_B T} {\bf t}\cdot {\bf z}_0-
V(\varphi)\\
\;& +{\Gamma\over R}\sin\varphi+
{\Gamma^2\over 16\ell_p}(t_x^2+t_y^2)] G=0,
\end{array}
\label{eq9}
\end{equation}
and the  force-extension and  torque-supercoiling  can
then be determined through the ground-state eigen relationsvalue
and eigenfunction of Eq. (\ref{eq9}). Finally, the
relation between extension and linking number is obtained
by elimination of torque $\Gamma$ from these two relations.  

\begin{figure}
\centerline{\psfig{file=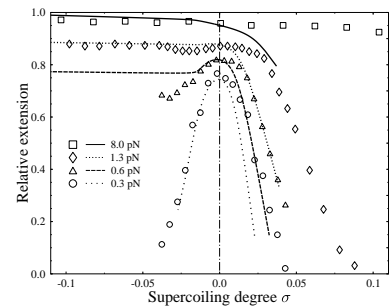,height=5cm}}
\vspace*{-1.0cm} 
\caption{
Extension vs   supercoiling relations
 at various stretching conditions.
The parameters for the theoretical curves are the
same as in Fig. \ref{fig1} and experimental data is
from Fig. 3 of [5].} \label{fig3}
\end{figure}

The numerically
 calculated relations between extension and
 supercoiling degree $\sigma$
at various fixed forces are shown in Fig. \ref{fig3} and
compared with the experiment of Strick {\it et al.}
\cite{strick96}. Here $\sigma$ is
defined by
$\sigma=(\langle Lk\rangle -\langle Lk\rangle_{\Gamma=0})/
\langle Lk\rangle_{\Gamma=0}$, 
where $\langle Lk\rangle_{\Gamma=0}=\int_0^L ds\langle
\sin\varphi\rangle_{\Gamma=0} /R$ is the
linking number for a torsionally relaxed DNA. 
The parameters for the theoretical curves in  Fig. \ref{fig3} are
the same as those of Figs. \ref{fig1} and \ref{fig2},   no adjustment 
 has ever been made to fit the data. For negatively supercoiled DNA, the theory is 
in  quantitative accordance
 with experiment (left half of Fig. \ref{fig3}).

\vspace*{1.0cm}

For $\sigma <0$, both theory and experiment give three
distinct regions of DNA elasticity:
(i) For forces $> 1.3$ pN, DNA extension does not 
shrink with the
increase of negative supercoiling, on the contrary, 
 it may even slightly increase as $|\sigma|$
increases.
(ii) For $1.3\geq f>0.3$ pN, there exists a critical negative supercoiling
degree $\sigma_c$. Extension of DNA shrinks as $\sigma$ decreases from $0$ to
$\sigma_c$, then it remains approximately constant as $\sigma$ further decreases.
 $\sigma_c\simeq -0.02$ at $0.6$ pN.
(iii) For $f\leq 0.3$ pN, DNA extension shrinks constantly with the increase of
$|\sigma|$. In this case,
 no evident  difference between  the behaviors of negatively 
and positively supercoiled DNAs is
 observed,  {\it i.e.}, DNA can be regarded
as achiral \cite{bouchiat98}.

Thus,   the complex elastical
property of a negatively supercoiled DNA 
as well as that of an overstretched DNA  can be
satisfactorily understood by the same framework.
In this context,  although DNA double-helix is 
 quite good at enduring external forces it is 
much weaker at  enduring  torques:
while   a force  $\sim$ $70$ pN is needed
for a
torsionally relaxed DNA 
to trigger  cooperative changes    of  configuration
\cite{cluzel96,smith96}, 
 $0.6$ pN is just sufficient for a  negatively 
supercoiled DNA with $\sigma$ as small as
$-2\%$.
This $``$shortcoming" of DNA  might have been
  well noticed and captured  
by various  proteins. For example, 
it seems that RecA protein stretches DNA
by exerting  a torque on the molecule \cite{subramanya96}.

However, as shown in the right half of Fig. \ref{fig3},
 for positively supercoiled DNA the agreement between
 theory and experiment is poor. It is possible that
positive supercoiling leads to  strong radial as well as
axial compressions on DNA basepair planes as to
make them shrink considerably or even corrupt.
In support of this point, 
recent experiment of Allemand {\it et al.} \cite{allemand98}
indicates that positively supercoiled DNA can
take on very surprising configurations with exposed  bases.
Therefore, it seems necessary  for us
to take into account the possible
deformability of DNA basepairs in our theory to understand
the elasticity of positively supercoiled DNA. 
We  plan to perform such an effort.

We wish to thank Prof. L.-S. Liu  
for his help on 
computer.  Z.H. is indebted to  
J.-Z. Lou  for informing him
a website of Fortran programs, to D. Thirumalai for
helpful correspondence, and  to  K. Kroy,
J. Yan  and Q.-H. Liu.


\begin{thebibliography}{100}
\bibitem[*]{zhouhj}  
To whom correspondence should be addressed.
\bibitem{remark1} Experimentally, the BPL of 
dsDNA is about $53$ nm, quite larer than that of a DNA single
strand \cite{smith96}.
\bibitem{remark4} In this work, we have  considered 
neither the possible
 tilting of the basepairs relative to the
central axis \cite{cluzel96}
 nor the possible nicks existed in DNA backbones
\cite{smith96}. It seems that
 although these effects can also 
be important in understanding DNA elasticity,  
they are not the dominating factors.
\bibitem{remark5} Because at low forces dsDNA
can be viewed as a wormlike chain with BPL of $53$ nm \cite{marko95a}.
\bibitem{remark6} These expressions for $r_0$ and $R$ follow from the facts that, 
for relaxed DNA, the axial distance between
 adjacent basepairs is $0.34$ nm and there are $10.5$ basepairs in 
 one period of the double-helix \cite{saenger84}.
\end{thebibliography}
\end{document}